\documentclass[preprint]{aastex61}

\newcommand\aastex{AAS\TeX}

\received{}
\revised{}
\accepted{}

\submitjournal{ApJL}

\shorttitle{\aastex\ An Inner Disk around DM Tau}
\shortauthors{Kudo et al.}

\begin{document}

\title{A Spatially Resolved AU-scale Inner Disk around DM Tau}

\correspondingauthor{Tomoyuki Kudo}
\email{kudotm@subaru.naoj.org}

\author{Tomoyuki Kudo}
\affiliation{SubaruTelescope, National Astronomical Observatory of Japan, 650 North A$'$ohoku Place, Hilo, HI 96720  U.S.A}

\author[0000-0002-3053-3575]{Jun Hashimoto}
\affiliation{Astrobiology Center, National Institutes of Natural Sciences, 2-21-1 Osawa, Mitaka, Tokyo 181-8588, Japan}

\author{Takayuki Muto}
\affiliation{Kogakuin University, 2665-1 Nakano, Hachioji, Tokyo 192-0015, Japan}

\author{Hauyu Baobab Liu}
\affiliation{European Southern Observatory, Karl-Schwarzschild-Str. 2, 85748 Garching, Germany}

\author{Ruobing Dong}
\affiliation{Department of Physics \& Astronomy, University of Victoria, Victoria, BC, V8P 1A1, Canada}

\author{Yasuhiro Hasegawa}
\affiliation{Jet Propulsion Laboratory, California Institute of Technology, Pasadena, CA 91109, U.S.A}

\author{Takashi Tsukagoshi}
\affiliation{National Astronomical Observatory of Japan, 2-21-1 Osawa, Mitaka, Tokyo 181-8588, Japan}

\author[0000-0003-0114-0542]{Mihoko Konishi}
\affiliation{Astrobiology Center, National Institutes of Natural Sciences, 2-21-1 Osawa, Mitaka, Tokyo 181-8588, Japan}



\begin{abstract}
We present Atacama Large Millimeter/submillimeter Array (ALMA) observations of the dust continuum emission at 1.3~mm and $^{12}$CO~$J = 2 \rightarrow 1$ line emission of the transitional disk around DM~Tau. DM~Tau's disk is thought to possess a dust-free inner cavity inside a few au, from the absence of near-infrared excess on its spectral energy distribution (SED). Previous submillimeter observations were, however, unable to detect the cavity; instead, a dust ring $\sim$20 au in radius was seen. The excellent angular resolution achieved in the new ALMA 
observations, 43$\times$31 mas, allows discovery of a 4~au radius inner dust ring, confirming previous SED modeling results. This inner ring is symmetric in continuum emission, but asymmetric in $^{12}$CO emission. The known (outer) dust ring at $\sim$20 au is recovered and shows azimuthal asymmetry with a strong-weak side contrast of $\sim$1.3. The gap between these two rings is depleted by a factor of $\sim$40 in dust emission relative to the outer ring. An extended outer dust disk is revealed, separated from the outer ring by another gap.  
The location of the inner ring is comparable to that of the main asteroid belt in the solar system. 
As a disk with a ``proto-asteroid belt,'' the DM Tau system offers valuable clues to disk evolution and planet formation in the terrestrial planet forming region.
\end{abstract}

\keywords{protoplanetary disks | stars: individual (DM Tau)}



\section{Introduction} \label{sec:intro}

An outstanding problem in planetesimal formation from aggregating dust in protoplanetary disks is radial drift of dust \citep{weid77,naka86}: 
particles embedded in a gaseous disk with surface density decreasing outwards feel a headwind, lose angular momentum to the gas, and drift towards the central star.
One solution to this problem is a dust trap \citep{rice06, johansen+09}, in which mm-sized particles are trapped and accumulate at a local gas pressure maximum. 
To facilitate the formation of planetesimals in protoplanetary disks on the scale of the inner solar system, dust traps at a few au from the star are needed. 
Such structures can be searched for by high angular resolution observations of mm continuum emission.

Many mechanisms have been proposed for producing pressure bumps in disks, such as the edges of gaps opened by planets (\citealt{zhu12, dong15}; note that one planet may produce multiple pressure bumps, \citealt{dong17a}). Magnetohydrodynamic (MHD) effects can also form pressure bumps in disks, generated by zonal flows \citep[e.g.,][]{johansen+09} or at the boundary of dead zones \citep[e.g.,][]{dzyurkevich10}. Pressure bumps may form at the locations of snowlines too, due to a change in the activity of the magnetorotational instability \citep[e.g.,][]{kretk07}. Dust trapped at radial pressure bumps appears to be annular rings in millimeter continuum observations. Such structures have been found in many objects, such as HL Tau \citep{almapart+15}, TW Hya \citep{andr16, tsukagoshi+16}, HD 163296 \citep{isel16}, and MWC 758 \citep{dong18}.

Our target, DM~Tau (SpT: M1; \citealp{keny95}, $T_{\rm eff}$: 3705~K; \citealp{andr11}, $M_{\rm *}$: 0.53~$M_{\odot}$; \citealp{piet07}, distance: 145~pc; \citealp{gaia18}), has a known transitional disk \citep{espaillat+14}. A central dust cavity $\sim$3~au in radius has been inferred based on its spectral energy distribution (SED) (\citealp{bergin04,calv05}, shown in Appendix). Previous SMA sub-millimeter continuum observations were not able to resolve the 3 au cavity due to insufficient angular resolution; instead, a dust ring at 19~au was discovered \citep{andr11}. Modeling of previous low resolution ALMA continuum observations (project ID:2013.1.00198.S; resolution $\sim$0\farcs4) suggested the presence of another faint dust ring at $\sim$80 au \citep{zhan16}. DM Tau has also been extensively studied in gas emission observations. \citet{bergin16a} resolved a C$_2$H emission ring at the edge of the dust continuum disk, and many other molecular species, such as H$_2$CO and CS, have been detected \citep{loomis15, semenov18}.

\section{Observations} \label{sec:obsres}
DM~Tau was observed with ALMA in band 6 in the C43--9 configuration on October 27, 2017, UT as part of the project 2017.1.01460.S, utilizing 47 antennas with the baseline length 
extending from 135.1~m to 14.9~km.
The observations were conducted in five spectral windows: two with bandwidths  117.188~MHz, velocity resolutions $\sim$0.166~km~s$^{-1}$, and
centered at 220.39868 GHz for $^{13}$CO~(2~$\rightarrow$~1) and 219.56035 GHz for C$^{18}$O~(2~$\rightarrow$~1); one with bandwidth 117.118~MHz, velocity resolution $\sim$0.079~km~s$^{-1}$, and 
centered at 230.53800 GHz for $^{12}$CO~(2~$\rightarrow$~1); and the last two windows for continuum observations with bandwidth 2.0~GHz. 
The precipitable water vapor was $\sim$0.5~mm during observations. The total on-source integration time was 66.1~min. The data were calibrated by the Common Astronomy Software Applications (CASA) package \citep{mcmu07} version 5.1.1, following the calibration scripts provided by ALMA. 
We had experimented with self-calibrating the new, high angular resolution ALMA data.
However, in these observations, only 16 antennas were located at $\gtrsim$2 km baselines.
With the 66.1 min on-source integration of our observations, our {\it uv} sampling at long baselines is insufficiently redundant.
In addition, the continuum flux of our target source is dominated by structures with relatively extended ($\sim$0.2$''$) angular scales.
As a result, the gain phase self-calibration flagged out over 50\% of the $\gtrsim$2 km baseline data even when using a per-scan solution interval and combining all spectral windows.
This is unfavorable for our major science case of resolving the innermost region of DM Tau.
On the other hand, the phase RMS of the $<$2 km baseline data is low, and the gain phase self-calibration does not help much.
Therefore, we decided not performing self-calibration.
We utilized the observations of the check (quasar) source J0449+1121 to assess how much our target source can be attenuated due to phase decoherence.
The continuum fluxes of J0449+1121 with and without phase self-calibration are 293 and 255 mJy, respectively.
This corresponds to a 13\% attenuation, which is much smaller than the  uncertainty of the dust mass opacity and the dust opacity depth in general.

We combined our data with another ALMA dataset \citep[project ID: 2013.1.00498.S; ][]{pini18} to recover the missing flux
($\sim$70 mJy out of a total of $\sim$110 mJy; \citealt{beck90}) 
due to the sparseness of short baseline data. The phase centers of long and short baseline data were determined separately by ellipse isophoto fitting at 10~$\sigma$ RMS noise in the dust continuum images synthesized by CASA with the \verb#CLEAN# task using a multi-scale multi-frequency deconvolution algorithm \citep{rau11}, and were shifted by \verb#fixvis# in the CASA tools. We compared the amplitudes as a function of $uv$-distance at less than 200~k$\lambda$ between our long baseline data and archival short baseline data, and confirmed their consistency. The CLEANed dust continuum image was synthesized with a Briggs weighting of 2.0 to maximize the signal-to-noise (S/N) ratio, providing RMS noise levels of 11~$\mu$Jy beam$^{-1}$. The total flux density after combining long and short baseline data is 116.75~$\pm$~0.14~mJy, consistent with previous single dish observations assuming a 10~\% uncertainty in absolute flux calibration. 

The $^{12}$CO~(2~$\rightarrow$~1) line data in both long and short baseline data were extracted by subtracting the continuum in the visibility space with  \verb#uvcontsub# in the CASA tools. The combined line cube was generated by the \verb#CLEAN# algorithm with a velocity resolution of 0.5~km/s, and was spatially smoothed with a circular Gaussian kernel of 75~mas by \verb#imsmooth# in CASA for presentation purposes. Though the $^{13}$CO and C$^{18}$O~(2~$\rightarrow$~1) line data were processed with the same procedure as $^{12}$CO, they were not detected significantly.

\section{Results} \label{sec:results}

\begin{figure}[tbh!]
\centering
\includegraphics[scale=0.4]{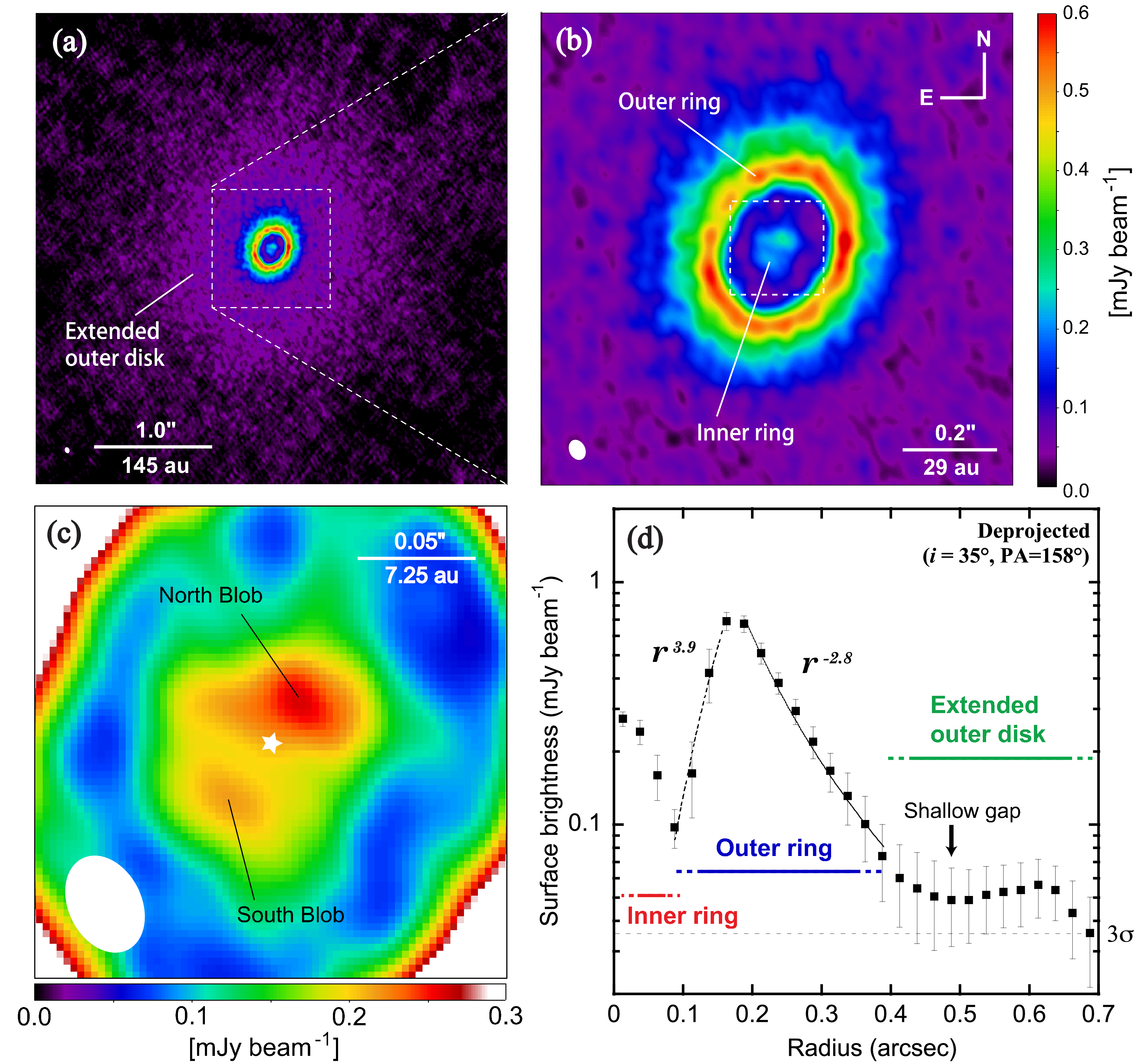} 
\caption{
1.3~mm dust continuum image around DM Tau. The synthesized beam size of 0.$''$043 $\times$ 0.$''$031 with a position angle of 22.7$^{\circ}$ is shown at the left bottom. The 1$\sigma$ noise level is 11~$\mu$Jy beam$^{-1}$.
(a): Overview of the DM~Tau disk. 
(b): Zoomed-up image to the outer ring. The color range is the same as in panel (a).
(c): Image focusing on the inner disk region indicated by a dashed white square in panel (b). The white star denotes the center of the outer dust ring determined by ellipse fitting (Table~1). 
(d): Azimuthally averaged radial profiles in the deprojected image. To minimize effects of the beam elongation and obtain a circular beam, we apply tapering in the CLEAN process of the deprojected image. The beam size is 0.$''$051 $\times$ 0.$''$050 and the noise level is 12~$\mu$Jy beam$^{-1}$. The 3$\sigma$ noise level of the deprojected image is indicated by the dashed line.
}
\label{fig:fig1}
\end{figure}

Figure~\ref{fig:fig1} shows the 1.3~mm dust continuum image of DM~Tau after the CLEANed process. We clearly resolve the dust disk into three components: an inner dust ring, an outer dust ring, and an extended outer disk (Figures~\ref{fig:fig1}a to c). The peak flux density at the inner and the outer rings is detected with 24 and 59~$\sigma$, respectively. An extended structure beyond $r=$0.$''$4 is also marginally detected with $\sim$5~$\sigma$. To derive the azimuthally averaged radial brightness profile (Figure~\ref{fig:fig1}d), we deprojected the dust continuum image in the visibility domain following \citet{zhan16}. The inclination and the position angle of the disk were derived by fitting an ellipse to the outer ring. The fitting results and derived disk's geometric parameters are shown in Table \ref{table:ellipse}. 

An inner dust ring at $r\sim$~0.$''$03 is discovered in our dust continuum images. The ring is spatially resolved
into a north and a south blobs (Figure~\ref{fig:fig1}c); the north blob is 20~$\pm$~8~\% (2.5~$\sigma$) brighter than the south one. More data are needed to confirm the apparent asymmetry.
The total flux density of the inner ring inside 0.$''$06 is 1.33~$\pm$~0.03~mJy. 
The contamination from possible free-free emission is less than 8\% (3$\sigma$ level), determined by extrapolating the flux density measured at 3.4~cm \citep{zapa17} to 1.3~mm assuming a spectral index of +0.6.
Assuming a distance of 145~pc, a dust opacity per gas mass $\kappa_{\nu}=$2.3~cm$^2$~g$^{-1}$ at 230~GHz \citep{beckwith+91}, a temperature of 100~K, and a gas-to-dust mass ratio of 100, the total mass of the inner disk is measured as 0.04~$M_{\rm Jup}$.

Our observations also clearly spatially resolved the outer dust ring and the gap between the two rings at 0.$''$18 and $r\sim$0.$''$1, respectively (Figures~\ref{fig:fig1}b and 1c). The dust continuum emission is detected with $\gtrsim$5~$\sigma$ in the gap region, suggesting that the gap is not dust free. The outer ring is asymmetric: the brightness contrast between the peak flux density at P.A.$=\sim$270$^{\circ}$ and that at the opposite position is 1.28~$\pm$~0.04. 
The inner edge of the outer ring is steeper ($I(r) \propto r^{3.9 \pm 0.3}$) than the outer edge ($I(r) \propto r^{-2.8 \pm 0.1}$). Because the outer ring is spatially resolved with a radial width of $\sim$0.$''$1 (15~au), the gradient difference is real. The extended structure beyond the outer ring has a nearly flat radial brightness profile (Figure~1d). A possible shallow gap at $r \sim$0.$''$5 can be seen in this structure as well.

\begin{figure}[tbh!]
\centering
\includegraphics[scale=0.35]{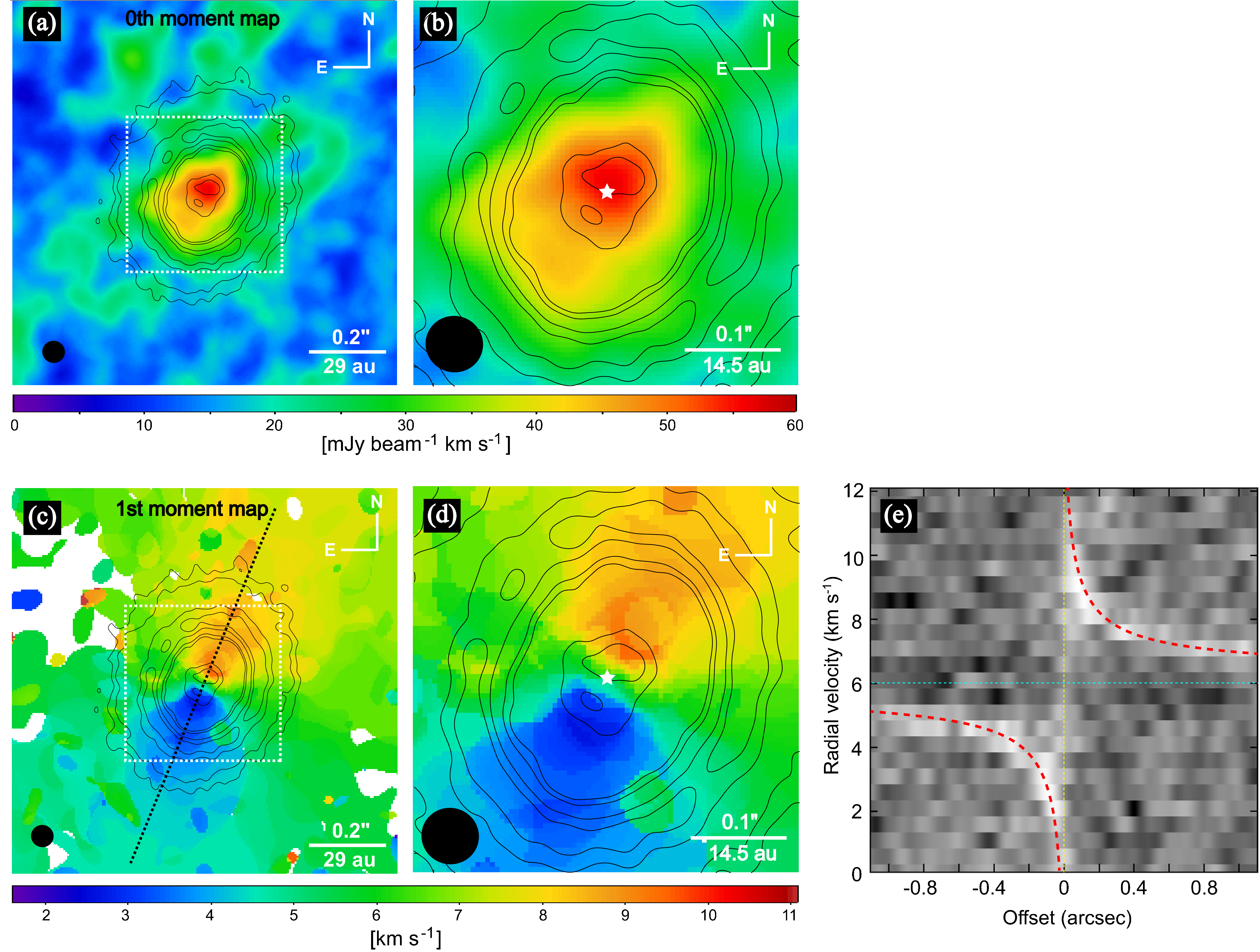} 
\caption{
CLEANed $^{12}$CO~(2~$\rightarrow$~1) line emission maps of the DM~Tau disk overlaying the dust continuum contours (14, 19, 34, and 44~$\sigma$). Panels~(a) and (b) show 0th moment maps, while panels~(c) and (d) show 1st moment ones. The $^{12}$CO images are spatially smoothed with a circular Gaussian of a $0.''075$ kernel. The white star denotes the center of the outer dust ring determined by ellipse fitting (Table~1). Panel~(e) shows the position-velocity diagram along the dashed line centered on the outer ring's center in panel~(c), and loci of peak emissions in the Keplerian disk around DM~Tau with a mass of $M =$0.53~$M_{\odot}$ and systemic velocity of 6.0~km s$^{-1}$ \citep{piet07}. 
}
\label{fig:fig2}
\end{figure}

Figures~\ref{fig:fig2}(a) and (b) show the integrated intensity (0th moment) map for $^{12}$CO obtained from 1.6 to 11.1~km~s$^{-1}$ with 1$\sigma$ = 3.5~mJy~beam$^{-1}$~km~s$^{-1}$. The peak emission is 56.5~mJy~beam$^{-1}$~km~s$^{-1}$, located at the north blob around the inner dust ring. The total integrated intensity is 185~$\pm$~7 Jy~km~s$^{-1}$ inside 0.$''$18. The intensity-weighted velocity (1st moment) map is shown in Figures \ref{fig:fig2}(c) and (d). The center of the CO gas motion nearly coincides with the center of the outer dust ring derived from the ellipse fitting, and might be closer to the north than to the south blob. To check whether the center of the outer ring is consistent with the rotational center of the CO gas, we plot loci of the peak emission of a Keplerian disk around a 0.53~$M_{\odot}$ star (DM~Tau's mass) in the position--velocity diagram (Figure~\ref{fig:fig2}e), finding that the outer ring's center is the center of the gas rotation.

\begin{deluxetable}{ccccc}
\tablecaption{Best-fit parameters of ellipse fitting in the outer ring\label{tab:table1}}
\tablehead{
\colhead{R.A. (ICRS)} & \colhead{Dec. (ICRS)} & \colhead{Radius}     & \colhead{$i$}          & \colhead{P.A.}\\
\colhead{}     & \colhead{}     & \colhead{($''$: au)} & \colhead{($^{\circ}$)} & \colhead{($^{\circ}$)}
}
\startdata
04:33:48.749 [0.001] & +18:10:09.64 [0.01] & 0.176$\pm$0.001: 25.5$\pm$0.2 & 35.2$\pm$0.7 & 157.8$\pm$1.0\\
\enddata
\tablecomments{
 Parentheses of R.A. and DEC. indicate 1$\sigma$ error in arcsecond. The heliocentric distance of the system (145~pc) is used to convert arcsecond into au.}
\label{table:ellipse}
\end{deluxetable}

\section{Model fitting} \label{sec:analysis}
To test whether the observed asymmetry in the outer dust ring is significant, we performed fitting for dust continuum emission in the visibility domain using a simple analytic disk model, and then subtracted the modeled disk from the data. Our disk model has a simple power-law radial profile with an exponential taper at the outside \citep[e.g.,][]{lynd74,hart98}: 
\begin{eqnarray*}
I(r) \propto \sum_{i=1}^{2} \alpha_{i} \left(\frac{r}{r_{c_{i}}}\right)^{-(q+\gamma_{i})}{\rm exp}\left[-\left(\frac{r}{r_{c_{i}}}\right)^{2-\gamma_{i}}\right], 
\end{eqnarray*}
where $\alpha_{i}$ and $r_{c_{i}}$ are a scaling factor and a characteristic scaling radius, respectively. 
The two global components ($i$) in the profile are: component~1 for the inner and outer dust rings, and component~2 for the extended outer disk (Figure~\ref{fig:fig3}a). The $q$ parameter is introduced to specify the radial dependence of the dust temperature $T_{d}$, that is $T_d \propto r^{-q}$. Optically thin emission in the Rayleigh--Jeans regime scales as
\begin{eqnarray*}
  I_{\nu} \propto B_{\nu}(T_{d})(1-e^{-\tau}) \propto T_{d}\tau, 
\end{eqnarray*}
where $B_{\nu}$ and $\tau$ are the blackbody intensity at frequency $\nu$ and the optical depth ($\tau = \kappa\Sigma$; where $\kappa$ and $\Sigma$ denote the opacity and the surface density, respectively).

In the radial direction, we have the following components with their scaling factors:

\begin{figure}[tbh!]
\includegraphics[scale=0.43]{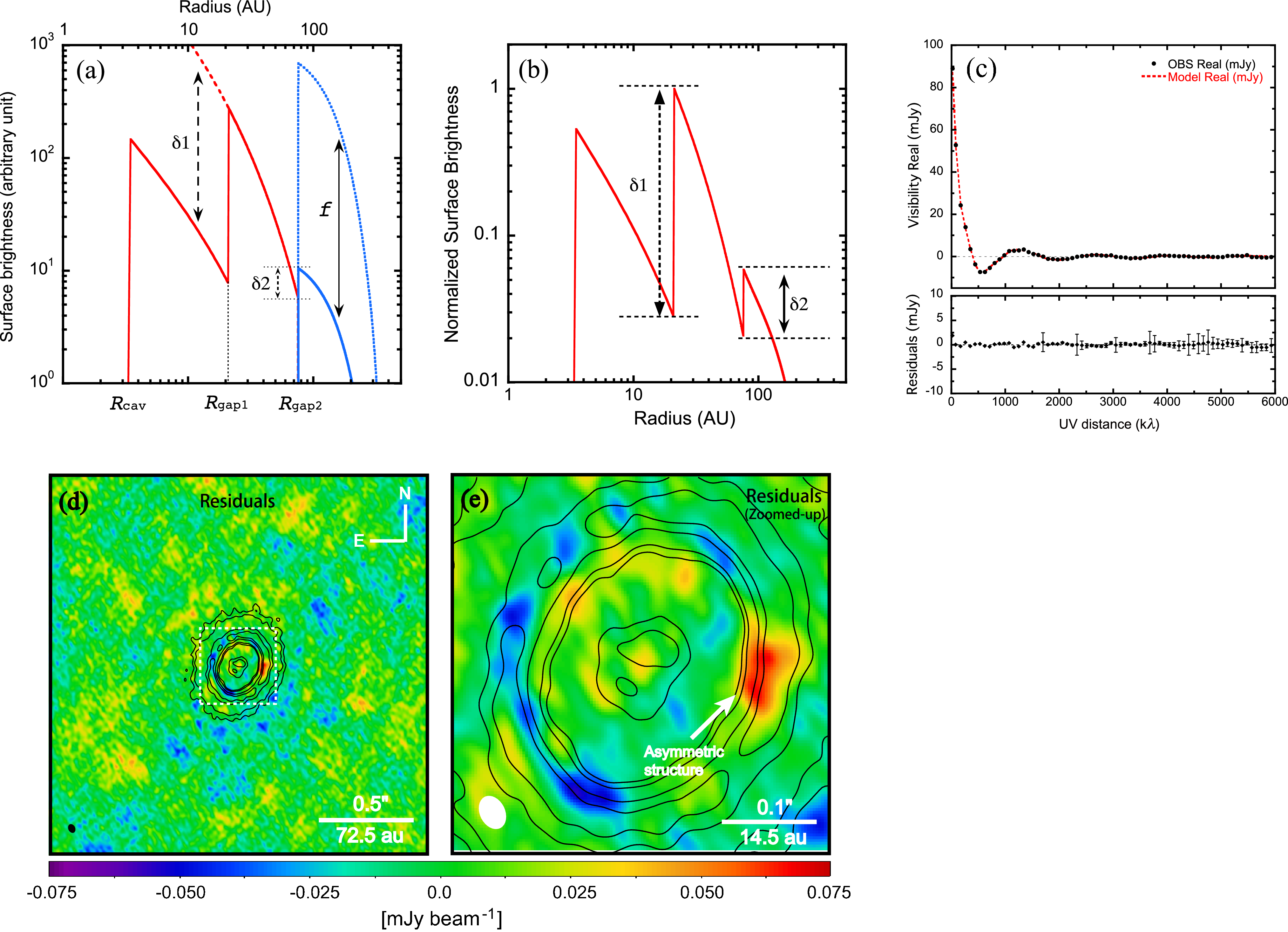}
\caption{
(a): Generic surface brightness model. The red and blue solid lines represent the surface brightness profiles of component~1 (the inner and outer rings at $r_{\rm cav}<r<r_{\rm gap_{1}}$ and $r_{\rm gap_{1}}<r<r_{\rm gap_{2}}$, respectively) and component~2 (the extended outer disk at $r_{\rm gap_{2}}<r$) scaled down by the scaling factor $\alpha_{i}$, respectively. At $r<r_{\rm cav}$, no dust emission is assumed due to absence of NIR excess in the DM~Tau's SED.
(b): Normalized surface brightness profile in our best-fit disk model. 
(c): Real part of the visibilities for the data (black dots) and the best-fit model (red line) in top panel. The bottom panel shows residual visibilities between best-fit and observations.
(d): Residual image (data$-$model) overlaying the dust continuum contours (14, 19, 34, and 44~$\sigma$). (e): Zoom-in version (see the white dotted box in panel (d)). 
}
\label{fig:fig3}
\end{figure}

\begin{eqnarray*}
\alpha_{1} &=& \left \{
\begin{array}{llllllll}
0                                        &{\rm for}&                 & & r  &<& r_{\rm cav}\\
\delta_{\rm {1}}                         &{\rm for}& r_{\rm cav}      &<& r  &<& r_{\rm gap_{1}}\\
1                                        &{\rm for}& r_{\rm gap_{1}} &<& r  &<& r_{\rm gap_{2}}\\
0                                        &{\rm for}& r_{\rm gap_{2}} &<& r, & &
\end{array}
\right. \\
\alpha_{2} &=& \left \{
\begin{array}{llllllll}
0                                        &{\rm for}&                 & & r  &<& r_{\rm gap_{2}}\\
f                                        &{\rm for}& r_{\rm gap_{2}} &<& r. & &
\end{array}
\right.
\end{eqnarray*} 

We normalize the total flux in the model ($F_{\rm total}$) to the observed value.
The disk inclination ($i$) and position angle (P.A.) are fixed as in Table 1. There are 11 free parameters in the model ($r_{\rm cav}$, $r_{\rm gap_{1}}$, $r_{\rm gap_{2}}$, $\delta_{1}$, $f$, $q$, $\gamma_{1}$, $\gamma_{2}$, $r_{c_{1}}$, $r_{c_{2}}$, and $F_{\rm total}$). The depletion factor $\delta_{2}$ at $r_{\rm gap_{2}}$ (Figure~\ref{fig:fig3}a) is measured after the calculations complete. To convert a modeled disk image to complex visibilities with identical $uv$-coverages of observations, we utilize the public python code {\sf vis\_sample}\footnote{{\sf vis\_sample} is publicly available at {\sf https://github.com/AstroChem/vis\_sample} or in the Anaconda Cloud at {\sf https://anaconda.org/rloomis/vis\_sample}}. The computed visibilities are deprojected in the $uv$-plane to calculate their azimuthal averages. For the fitting, we used the Markov Chain Monte Carlo (MCMC) method in the {\sf emcee} package \citep{foreman-mackey+2013}. Our calculations used flat priors with the parameter ranges summarized in Table~\ref{tab:table2}. The burn-in phase (from initial conditions to reasonable sampling) employs 500 steps, and we run another 500 steps for convergence, totaling 1000 steps with 100 walkers. 

The fitting result and the best-fit surface brightness profile are shown in Table~\ref{tab:table2} and Figure~\ref{fig:fig3}(b), respectively. The corner plot with the MCMC posteriors is also shown in Appendix. The depletion factor $\delta_{2}$ at $r_{\rm gap_{2}}$ is measured as $\sim$0.356 using the best-fit brightness profile in Figure~\ref{fig:fig3}(b). The reduced-$\chi^{2}$ calculated with the observed and modeled visibilities in Figure~\ref{fig:fig3}(c) is 2.4. 
The residual map prepared using the residual visibilities (data$-$model) is shown in Figures~\ref{fig:fig3}(d) and (e). 
The residual map shows a structure at $\sim$7~$\sigma$ level to the west in the outer dust ring, indicating a real azimuthal asymmetry.

\begin{deluxetable}{ccccccccccc}
\tabletypesize{\tiny}
\tablecaption{Results of MCMC fitting and its parameter ranges \label{tab:table2}}
\tablehead{
\colhead{$R_{\rm cav}$} & \colhead{$R_{\rm gap_{1}}$} & \colhead{$R_{\rm gap_{2}}$} & \colhead{$\delta_{\rm 1}$} & \colhead{$\delta_{\rm 2}$} & \colhead{$q$} & \colhead{$\gamma$1} & \colhead{$\gamma$2} & \colhead{$R_{\rm c1}$} & \colhead{$R_{\rm c2}$} & \colhead{Flux} \\
\colhead{(au)}              & \colhead{(au)}              & \colhead{(au)}          & \colhead{}          & \colhead{}          & \colhead{}    & \colhead{}          & \colhead{}          & \colhead{(au)}         & \colhead{(au)}         & \colhead{{mJy}} 
}
\startdata
 3.16$^{+0.22}_{-0.23}$ &  21.00$^{+0.02}_{-0.02}$ &  75.64$^{+0.37}_{-0.39}$ & 0.028$^{+0.001}_{-0.001}$ & $\sim$0.356 & 0.01$^{+0.01}_{-0.02}$ & 1.10$^{+0.05}_{-0.03}$ &  0.01$^{+0.02}_{-0.01}$ & 18.28$^{+0.90}_{-1.14}$ & 124.10$^{+1.05}_{-1.05}$ & 93.30$^{+0.52}_{-0.46}$ \\
\{0.00-7.25\} & \{14.50-29.00\} & \{72.50-101.50\} & \{1.000-0.001\} & --- & \{0.00-1.00\} & \{0.00-2.00\} & \{0.00-2.00\} & \{0.00-29.00\} & \{58.00-145.00\} & \{90.0-105.0\} \\
\enddata
\tablecomments{
 Parentheses describe parameter ranges in our MCMC calculations. The errors in $\gamma_{i}$ are large due to local maxima in calculations.
 The depletion factor of $\delta_{\rm 2}$ is measured in Figure~3(b). The factor of log~$f$ is $-1.82^{+0.05}_{-0.07}$. 
 }
\end{deluxetable}

\section{Discussions} \label{sec:discus}
The most intriguing result in our continuum observations is the detection of the inner dust ring at $r\sim$4~au and the cavity inside. 
Combining the SED and measured accretion rate of the system ($\dot{M}~\sim 6 \times $10$^{-9}$~$M_{\odot}$/yr; \citealp{mana14}, 
only slightly lower than that of typical T~Tauri~stars; \citealp{naji15}), we now have a more complete picture of its inner region: the cavity inside $r\sim$4~au has no detectable dust, consistent with the absence of NIR excess on the SED; however the cavity must have a substantial amount of gas in order to sustain a close-to-normal accretion rate.
While dust cavities are commonly found in ALMA continuum observations nowdays \citep[e.g.][see also \citealt{vandermarel18} for a gallery]{hashimoto+15, isel16, tang17, dong17b}, the inner cavity in the DM Tau disk, together with the cavity at 2.4 au in the TW Hya disk \citep{andr16, tsukagoshi+16}, are among the smallest, visible only in long baseline ALMA observations.

The origin of the inner cavity is unclear. 
The measured close-to-normal accretion rate of DM Tau disfavors photoevaporation \citep[e.g.,][]{alexander06}. 
A planet can open a gap in gas (e.g., Lin \& Papaloizo 1993), in which case the outer edge of the gap (even a shallow on) acts as a ``dust filter'', trapping $\sim$millimeter-sized large dust grains and forming a cavity in them \citep[e.g.,][]{rice06, zhu12}, consistent with our ALMA dust observations. Note that our gas observations are performed with $^{12}$CO, to which the disk easily becomes optically thick. We therefore cannot detect a possible gas gap inside the edge of the inner dust cavity. Future observations of optically thinner CO isotopologues are needed to probe the gas surface density structure across the inner cavity \citep[e.g.,][]{vandermarel16}.
However, as argued in \citet{zhu11}, disk--planet interactions have difficulties in depleting the small ($\micron$-sized) dust in the inner disk, which tend to flow in with the gas.
The DM Tau disk is at the extreme --- there is no detectable dust inside the inner cavity based on the SED, while CO emission extends all the way toward the central star.  

Alternatively, the presence of gas and the absence of small dust inside the cavity may be explained as icy dust being evaporated inside the cavity, fully or partially replenishing the gas. To produce this scenario, the dust that enters the inner cavity has to comprise fully evaporable volatiles (e.g., water), and the gas inside the cavity must be rich in them.
If all small icy grains do not evaporate, however, the decreased grain size inside the evaporation front increases the fragmentation efficiency \citep[e.g.,][]{pini17}, thus enriching the cavity with small grains well coupled to the gas, inconsistent with the SED.
Overall, both the planet scenario and the grain evaporation scenario have advantage and disadvantage, and additional observations are needed to determine the origin of the inner cavity.

The inner ring is located at a strikingly similar distance to the main asteroid belt in the solar system between Mars and Jupiter. The main belt, located beyond the water snowline at 2.7 AU \citep[e.g.,][]{abe00}, is thought to have profoundly impacted water delivery onto Earth \citep[e.g.,][]{morb00}. A fraction of the water snowballs in the main belt found their way into the inner solar system and bombarded the early Earth, as their orbits were perturbed by Jupiter. The water snowline in the DM Tau system, with a host star less luminous than the proto-Sun, should be located closer to its host star than that in the solar system \citep[see also][]{mart13, notsu16}. Should terrestrial planets be forming inside its snowline, water delivery from the inner ring onto the inner planets might be plausible.

As shown in \S~\ref{sec:analysis}, an axisymmetric inner dust ring is consistent with the data.
On the other hand, $^{12}$CO line emission is clearly non-axisymmetric on an $\sim$au scale --- the peak $^{12}$CO emission is located at the north blob (Figure~\ref{fig:fig2}), while the rotational center of CO roughly coincides with the stellar location and the center of the inner cavity.
A candidate massive planet with $M_{p} \sim$3~$M_{\rm Jup}$ \citep[COND model at 1~Myr;][]{baraffe+03} at a separation of $r\sim$6~au was detected by Keck sparse aperture masking interferometric observations \citep{will16}. Whether this candidate planet introduces the asymmetry seen in CO emission is yet to be explored.

The continuum emission in the gap between the inner and outer rings ($\sim$4--20 au) is substantially suppressed by a factor of $\sim$40 relative to the outer ring. The gap also has a steep inner edge (Figure~1d). Both features are consistent with predictions of gap opening by planets \citep{zhu11, dong15}, but inconsistent with some other gap formation mechanisms such as the secular gravitational instability \citep[SGI; e.g.,][]{taka14} and sintering of dust grains \citep{okuzumi+16} --- the latter is expected to produce shallow gaps with depletion factors less than 10 \citep{okuzumi+16}, while the formation timescale of the ring at $r \sim$10--20~au in the former mechanism is much longer than DM~Tau's age.

In contrast to the symmetric inner ring, the outer ring has a small azimuthal asymmetry at P.A. $\sim$270$^{\circ}$, similar to the asymmetries found in a few other disks, but with one of the lowest contrast levels, $\sim$1.3:1 (cf., $\sim$130 in Oph~IRS~48; \citealt{vandermarel13}, $\sim$30 in HD~142527; \citealt{fukagawa+13}).
This asymmetry may be caused by a local enhancement in the dust surface density, possibly the remnants of particle trapping in a vortex~\citep{barge17}, or temperature, if the south-west side is the far side while we observe the inner rim of the outer ring there \citep[see e.g., ][for a discussion of HD~142527]{muto+15,soon+17}.  
Note that if the outer ring is optically thick and the asymmetry traces dust surface density variations, the small contrast seen in the surface brightness may not trace the contrast in the dust surface density, which can be much bigger than 1.3.
Our observations also reveal an extended outer disk beyond the outer ring at $r>$0\farcs4 in the dust continuum, and a shallow gap inside. The presence of these structures has been previously suggested by \citet{zhan16} and \citet{bergin16b}.

Our new ALMA observations reveal exciting new details, and yet raise more questions, in the DM~Tau system. With an exo-asteroid belt under formation, the DM~Tau disk will continue to offer crucial insights into disk evolution and planet formation in the terrestrial--planet--forming region.

\acknowledgments
We thanks an anonymous referee for a helpful review of the manuscript.
This paper makes use of the following ALMA data: ADS/JAO.ALMA\#2017.1.01460.S and ADS/JAO.ALMA\#2013.1.00498.S. ALMA is a partnership of ESO (representing its member states), NSF (USA) and NINS (Japan), together with NRC (Canada), NSC and ASIAA (Taiwan), and KASI (Republic of Korea), in cooperation with the Republic of Chile. The Joint ALMA Observatory is operated by ESO, AUI/NRAO and NAOJ. This work was supported by NAOJ ALMA Scientific Research Grant Number 2016-02A and JSPS KAKENHI Grant Numbers 17K14258, 17K14244, 26800106, 15H02074 and 17H01103. Y.H is supported by JPL/Caltech under a contract from NASA.

\facility{ALMA} 

\software{CASA v5.1.1; \citep{mcmu07}, \sf vis\_sample \rm(https://github.com/AstroChem/vis\_sample), \sf emcee \rm \citep{foreman-mackey+2013} }

\clearpage
\appendix

In this section, we present the DM Tau's SED and the corner plot with the MCMC posteriors in Figure~\ref{fig:figA1} calculated in \S~\ref{sec:analysis}.

\begin{figure}[tbh!]
\includegraphics[scale=1.0]{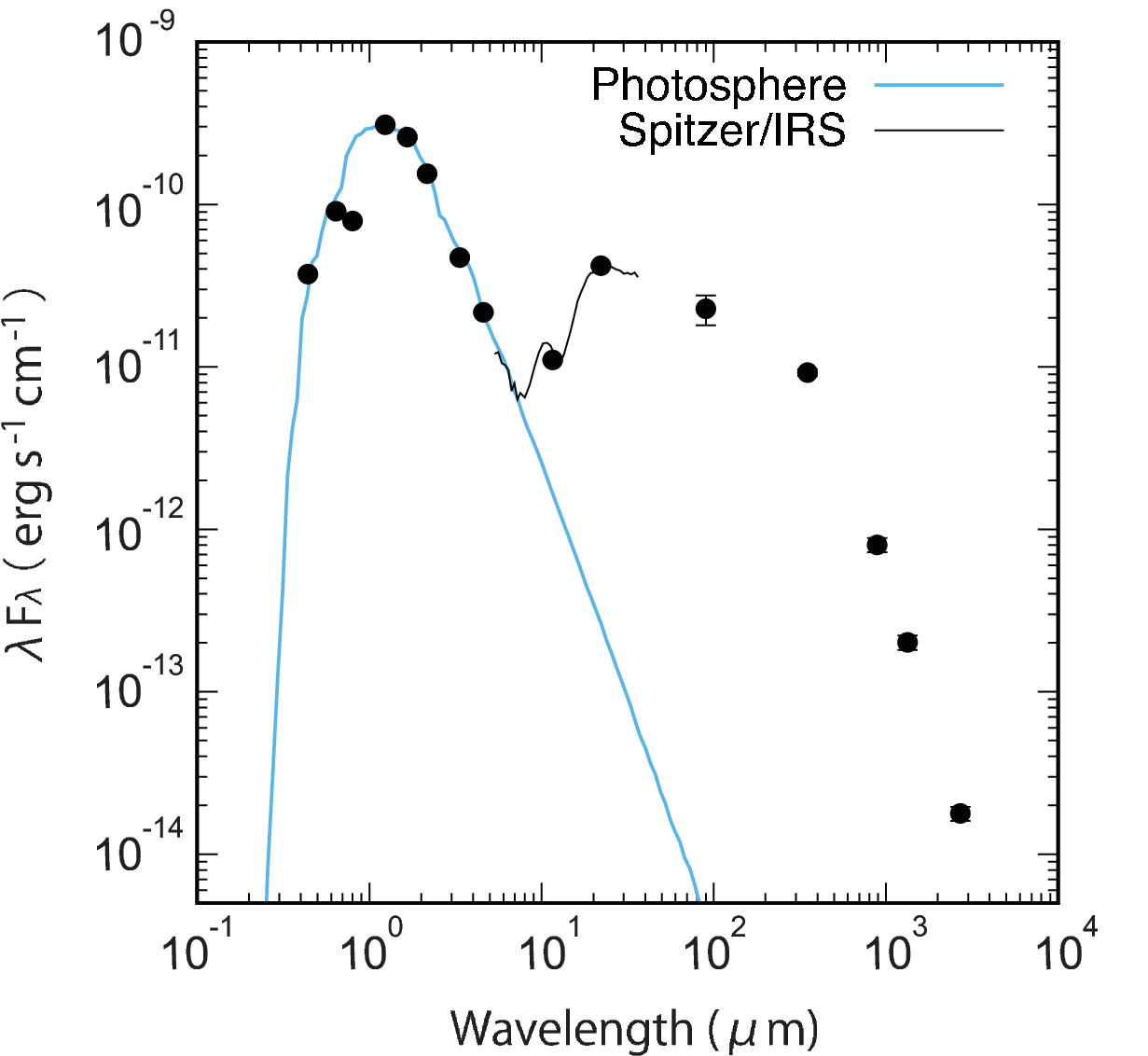}
\caption{
DM Tau's SED
}
\label{fig:DMTauSED}
\end{figure}

\begin{figure}[tbh!]
\includegraphics[scale=0.35]{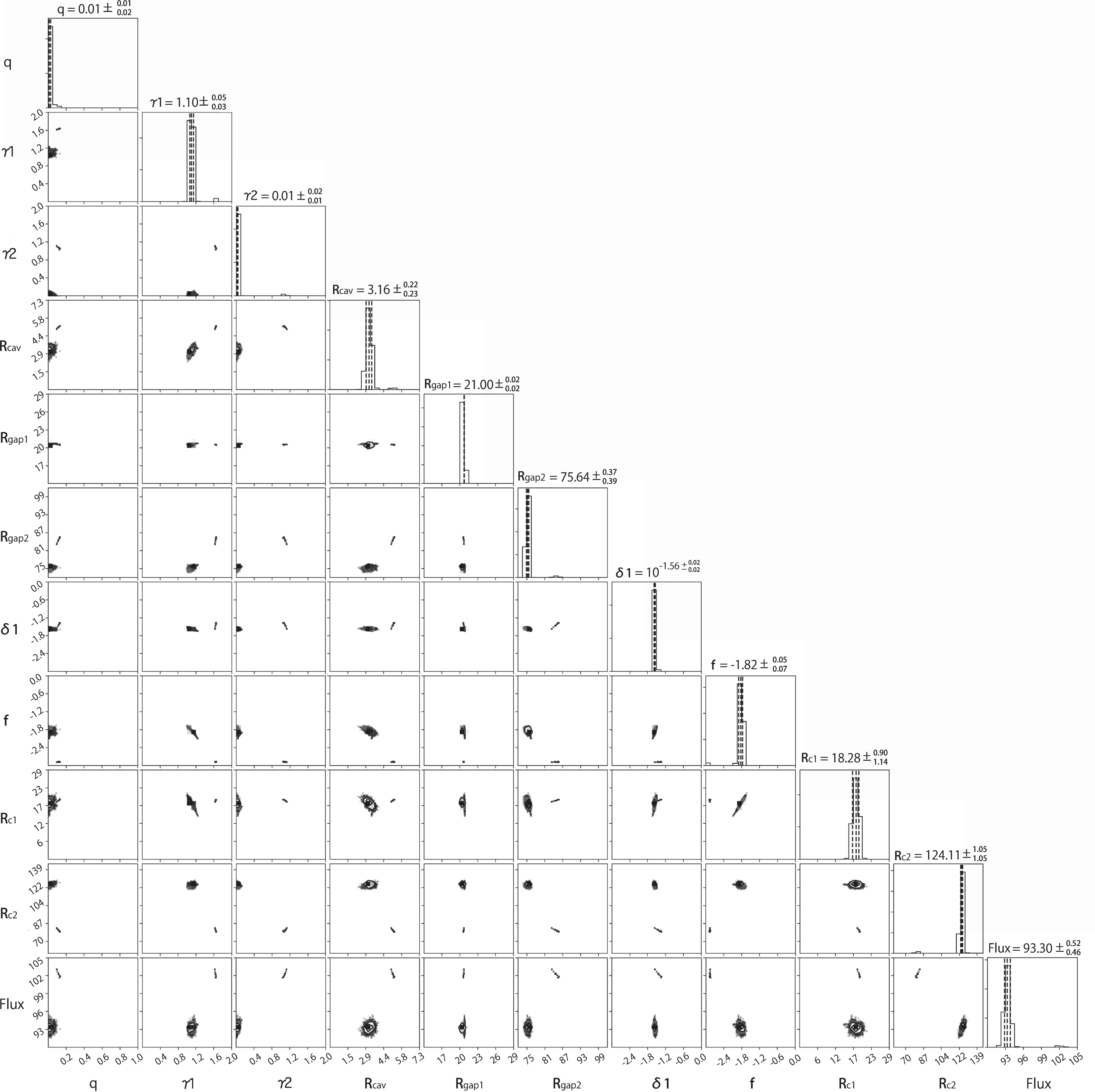}
\caption{
Corner plot with the MCMC posterior probability distribution calculated in \S~\ref{sec:analysis}. The histograms on diagonal are marginal distributions of 11 parameters, provided by the last 500 steps of the 100 walkers' chain. The vertical dashed lines in the histograms represent the median values and the 1~$\sigma$ confidence intervals of parameters, which are also shown in the titles. The off-diagonal plots show the correlation for corresponding pairs of parameters.
}
\label{fig:figA1}
\end{figure}


\clearpage

\begin{thebibliography}{}
\bibitem[Abe et al. (2000)]{abe00}Abe, Y., Ohtani, E., Okuchi, T., Righter, K., \& Drake, M. 2000, Water in the Early Earth, ed. R. M. Canup, K. Righter, \& et al., 413-433
\bibitem[Alexander et al. (2006)]{alexander06}Alexander, R. D., Clarke, C. J., \& Pringle, J. E. 2006, \mnras, 369, 229
\bibitem[ALMA Partnership et al. (2015)]{almapart+15} ALMA Partnership, Brogan, C. L., P\'erez, L. M., et al. 2015, \apjl, 808, L3
\bibitem[Andrews et al. (2011)]{andr11}Andrews, S. M., Wilner, D. J., Espaillat, C., et al. 2011, \apj, 732, 42
\bibitem[Andrews et al. (2016)]{andr16}Andrews, S. M., Wilner, D. J., Zhu, Z., et al. 2016, \apjl, 820, L40
\bibitem[Baraffe et~al.(2003)]{baraffe+03} Baraffe, I., Chabrier, G., Barman, T.~S., Allard, F., \& Hauschildt, P.~H. 2003, \aap, 402, 701
\bibitem[Barge et al. (2017)]{barge17}Barge, P., Ricci, L., Carilli, C. L., \& Previn-Ratnasingam, R. 2017, \aa, 605, A122
\bibitem[Beckwith \& Sargent (1991)]{beckwith+91} Beckwith, S. V. W., \& Sargent, A. I. 1991, \apj, 381, 250
\bibitem[Beckwith et al. (1990)]{beck90} Beckwith, S. V. W., \& Sargent, A. I. 1991, \apj, 381, 250
\bibitem[Bergin et al. (2004)]{bergin04} Bergin, E., Calvet, N., Sitko, M. L., et al. 2004, \apjl, 614, L133
\bibitem[Bergin et al. (2016a)]{bergin16a} Bergin, E. A., Du, F., Cleeves, L. I., et al. 2016a, \apj, 831, 101
\bibitem[Bergin et al. (2016b)]{bergin16b} ---. 2016b, \apj, 831, 101
\bibitem[Calvet et al. (2005)]{calv05}Calvet, N., D'Alessio, P., Watson, D. M., et al. 2005, \apjl, 630, L185
\bibitem[Dong et al. (2017a)]{dong17a} Dong, R., Li, S., Chiang, E., \& Li, H. 2017a, \apj, 843, 127
\bibitem[Dong et al. (2015)]{dong15}Dong, R., Zhu, Z., \& Whitney, B. 2015, \apj, 809, 93
\bibitem[Dong et al. (2017b)]{dong17b}Dong, R., van der Marel, N., Hashimoto, J., et al. 2017b, \apj, 836, 201
\bibitem[Dong et al. (2018)]{dong18} Dong, R., Liu, S.-y., Eisner, J., et al. 2018, \apj, 860, 124
\bibitem[Dzyurkevich et al. (2010)]{dzyurkevich10} Dzyurkevich, N., Flock, M., Turner, N. J., Klahr, H., \& Henning, T. 2010, \aa, 515, A70
\bibitem[Espaillat et~al.(2014)]{espaillat+14} Espaillat, C., Muzerolle, J., Najita, J., et~al. 2014, Protostars and Planets VI, 497
\bibitem[Foreman-Mackey et~al.(2013)]{foreman-mackey+2013} Foreman-Mackey, D., Hogg, D.~W., Lang, D., \& Goodman, J. 2013, \pasp, 125, 306
\bibitem[Fukagawa et al. (2013)]{fukagawa+13}Fukagawa, M., Tsukagoshi, T., Momose, M., et al. 2013, \pasj, 65, L14
\bibitem[Gaia Collaboration et al. (2018)]{gaia18} Gaia Collaboration, Brown, A. G. A., Vallenari, A., et al. 2018, ArXiv e-prints, arXiv:1804.09365
\bibitem[Hartmann et~al.(1998)]{hart98} Hartmann, L., Calvet, N., Gullbring, E., \& D'Alessio, P. 1998, \apj, 495, 385
\bibitem[Hashimoto et al. (2015)]{hashimoto+15}Hashimoto, J., Tsukagoshi, T., Brown, J. M., et al. 2015, \apj, 799, 43
\bibitem[Isella et~al.(2016)]{isel16} Isella, A., Guidi, G., Testi, L., et~al. 2016, Phys. Rev. Lett., 117, 251101
\bibitem[Johansen et al. (2009)]{johansen+09}Johansen, A., Youdin, A., \& Klahr, H. 2009, \apj, 697, 1269
\bibitem[Kenyon \& Hartmann (1995)]{keny95} Kenyon, S. J., \& Hartmann, L. 1995, \apjs, 101, 117
\bibitem[Kretke \& Lin (2007)]{kretk07}  Kretke, K. A., \& Lin, D. N. C. 2007, \apjl, 664, L55
\bibitem[Loomis et al. (2015)]{loomis15}  Loomis, R. A., Cleeves, L. I., \"Oberg, K. I., Guzman, V. V., \& Andrews, S. M. 2015, \apjl, 809, L25
\bibitem[Lynden-Bell \& Pringle(1974)]{lynd74} Lynden-Bell, D., \& Pringle, J.~E. 1974, \mnras, 168, 603
\bibitem[Manara et al. (2014)]{mana14}Manara, C. F., Testi, L., Natta, A., et al. 2014, \aa, 568, A18
\bibitem[Martin \& Livio (2013)]{mart13}Martin, R. G., \& Livio, M. 2013, \mnras, 428, L11
\bibitem[McMullin et~al.(2007)]{mcmu07} McMullin, J.~P., Waters, B., Schiebel, D., Young, W., \& Golap, K. 2007, in Astronomical Society of the Pacific Conference Series, Vol. 376, Astronomical Data Analysis Software and Systems XVI, ed. R. A. Shaw, F. Hill, \& D. J. Bell, 127
\bibitem[Morbidelli et al. (2000)]{morb00}Morbidelli, A., Chambers, J., Lunine, J. I., et al. 2000, Meteoritics and Planetary Science, 35, 1309
\bibitem[Muto et al. (2015)]{muto+15}Muto, T., Tsukagoshi, T., Momose, M., et al. 2015, \pasj, 67, 122
\bibitem[Najita et al. (2015)]{naji15} Najita, J. R., Andrews, S. M., \& Muzerolle, J. 2015, \mnras, 450, 3559
\bibitem[Nakagawa et al. (1986)]{naka86}Nakagawa, Y., Sekiya, M., \& Hayashi, C. 1986, \icarus, 67, 375
\bibitem[Notsu et al. (2016)]{notsu16}Notsu, S., Nomura, H., Ishimoto, D., et al. 2016, \apj, 827, 113
\bibitem[Okuzumi et~al.(2016)]{okuzumi+16} Okuzumi, S., Momose, M., Sirono, S.-i., Kobayashi, H., \& Tanaka, H. 2016, \apj, 821, 82
\bibitem[Pi\'etu et al. (2007)]{piet07}Pi\'etu, V., Dutrey, A., \& Guilloteau, S. 2007, \aa, 467, 163
\bibitem[Pinilla et al. (2017)]{pini17} Pinilla, P., Pohl, A., Stammler, S. M., \& Birnstiel, T. 2017, \apj, 845, 68
\bibitem[Pinilla et al. (2018)]{pini18} Pinilla, P., Tazzari, M., Pascucci, I., et al. 2018, ArXiv e-prints, arXiv:1804.07301
\bibitem[Rau \& Cornwell (2011)]{rau11} Rau, U., \& Cornwell, T. J. 2011, \aa, 532, A71
\bibitem[Rice et al. (2006)]{rice06}Rice, W. K. M., Armitage, P. J., Wood, K., \& Lodato, G. 2006, \mnras, 373, 1619
\bibitem[Semenov et al. (2018)]{semenov18}  Semenov, D., Favre, C., Fedele, D., et al. 2018, ArXiv e-prints, arXiv:1806.07707
\bibitem[Soon et al. (2017)]{soon+17}Soon, K.-L., Hanawa, T., Muto, T., Tsukagoshi, T., \& Momose, M. 2017, \pasj, 69, 34
\bibitem[Takahashi \& Inutsuka(2014)]{taka14} Takahashi, S.~Z., \& Inutsuka, S.-i. 2014, \apj, 794, 55
\bibitem[Tang et al. (2017)]{tang17}Tang, Y.-W., Guilloteau, S., Dutrey, A., et al. 2017, \apj, 840, 32
\bibitem[Tsukagoshi et al. (2016)]{tsukagoshi+16}Tsukagoshi, T., Nomura, H., Muto, T., et al. 2016, \apjl, 829, L35
\bibitem[van der Marel et al. (2016)]{vandermarel16}van der Marel, N., Cazzoletti, P., Pinilla, P., \& Garu, A. 2016, \apj, 832, 178
\bibitem[van der Marel et al. (2013)]{vandermarel13}van der Marel, N., van Dishoeck, E. F., Bruderer, S., et al. 2013, Science, 340, 1199
\bibitem[van der Marel et al. (2018)]{vandermarel18}van der Marel, N., Williams, J. P., Ansdell, M., et al. 2018, \apj, 854, 177
\bibitem[Weidenschilling. (1977)]{weid77} Weidenschilling, S. J. 1977, \mnras, 180, 57
\bibitem[Willson et al. (2016)]{will16}Willson, M., Kraus, S., Kluska, J., et al. 2016, \aa, 595, A9
\bibitem[Zapata et al.(2017)]{zapa17} Zapata, L. A., Rodr\'iguez, L. F., \& Palau, A. 2017, \apj, 834, 138
\bibitem[Zhang et al. (2016)]{zhan16} Zhang, K., Bergin, E. A., Blake, G. A., et al. 2016, \apjl, 818, L16
\bibitem[Zhu et al. (2012)]{zhu12}Zhu, Z., Nelson, R. P., Dong, R., Espaillat, C., \& Hartmann, L. 2012, \apj, 755, 6
\bibitem[Zhu et al. (2011)]{zhu11}Zhu, Z., Nelson, R. P., Hartmann, L., Espaillat, C., \& Calvet, N. 2011, \apj, 729, 47
\end{thebibliography}

\end{document}